\documentclass{article}
\usepackage{etex}
\reserveinserts{28}
\usepackage{setspace, amsmath, amssymb, amsthm}
\usepackage[colorlinks = true]{hyperref}
\usepackage{xcolor}
\usepackage[margin = 2.5cm]{geometry}
\usepackage[english]{babel}
\usepackage[utf8x]{inputenc}
\usepackage{amsfonts,tikz}
\usepackage{comment}
\usepackage{graphicx} 
\usepackage{epsfig}
\usepackage{epstopdf}
\usepackage{subcaption}
\usetikzlibrary{calc}
\usepackage{csquotes}
\usepackage{caption}
\usepackage{array}
\usetikzlibrary{fit}
\usepackage{tikz-cd}
\usetikzlibrary{arrows}
\usetikzlibrary{positioning}
\newtheorem{theorem}{Theorem}[section]

\newtheorem{corollary}{Corollary}[theorem]

\newtheorem{mydef-remark}{Remark}
\tikzstyle{vertex}=[circle, draw, inner sep=0pt, minimum size=4pt]
\label{marker}
\newcommand{\vertex}{\node[vertex]}

\newcommand{\C}{{\mathbb{C}}}

\newcommand{\ket}[1]{|{#1}\rangle}

\newcommand{\abs}[1]{\lvert #1 \lvert }

\title{A Graph-Theoretical Framework to Analyse Zero Discord Quantum States} 

\author{Anoopa Joshi \thanks{
Email: \texttt{PG201283001@iitj.ac.in}} \\Indian Institute of Technology Jodhpur, Rajasthan, India\\Parvinder Singh \thanks{
Email: \texttt{parvinder.singh@cup.edu.in}} \\Central University of Punjab, Bathinda, India\\Atul Kumar \thanks{
Email: \texttt{atulk@iitj.ac.in}} \\ Indian Institute of Technology Jodhpur, Rajasthan, India  }

\date{}
\begin{document}
\maketitle
\begin{abstract}
		This article comprehensively explores matrices and their prerequisites for achieving positive semi-definiteness. The study delves into a series of theorems concerning pure quantum states in the context of weighted graphs. The main objective of this study is to establish a graph-theoretic framework for the study of quantum discord and to identify the necessary and sufficient conditions for zero quantum discord states using unitary operators. This research aims to advance the understanding of quantum discord and its implications for quantum information theory with a graph-theoretic framework.

\end{abstract}
	
	\section{Introduction}

	In the bipartite and multipartite quantum systems, the phenomenon of entanglement exhibits that a local measurement on one subsystem will have a profound impact on the other subsystem due to the existence of nonlocal correlation \cite{nielsen2000quantum,dur2003multiparticle,horodecki2001separability,barnett2009quantum,bennett1996concentrating,wootters1998entanglement,wootters2001entanglement}. However, it is worth noting that entanglement is not always a prerequisite for demonstrating nonlocal characteristics in quantum systems \cite{bennett1999quantum,bennett1996mixed}. For example, a quantum computing framework- deterministic quantum computation using single qubit- utilizing highly mixed states was proposed in 1998  \cite{knill1998power,datta2008quantum}. Notably, this model was successfully implemented experimentally in 2008 \cite{lanyon2008experimental}, and serves as an excellent illustration that certain highly mixed states, even when separable, exhibit intrinsic non-classical correlations which hold promising potential for applications in quantum computing. Moreover, investigations have revealed that quantum correlation exhibits greater resilience than entanglement in the presence of noise, rendering quantum algorithms relying solely on nonlocal correlations more robust compared to those relying on entanglement \cite{werlang2009robustness,wang1993some,wang2010non,fanchini2010non}.
		
	\par Quantum discord is a measure of non-classical quantum correlation, and it has been extensively studied for achieving efficient, secure, and optimal quantum communication beyond the scope of entanglement. Quantifying quantum correlation in a bipartite state is achieved by measuring quantum discord, initially introduced by Ollivier and Zurek \cite{ollivier2001quantum}. Obtaining an analytical expression of quantum discord for arbitrary two-qubit states is challenging due to the complex optimization over local measurements. Hence, determining whether a quantum state has zero quantum discord is crucial in distinguishing its quantum characteristics. For instance, it has been established that a vanishing quantum discord between the principal system and its surroundings is necessary and sufficient for characterizing the system's dynamics through a completely positive map \cite{shabani2009vanishing,luo2016entanglement,dakic2010necessary,modi2012classical,rulli2011global}. Recently, researchers proposed a necessary and sufficient condition for nonzero quantum discord using singular value decomposition of a correlation matrix \cite{dakic2010necessary,deutsch1996quantum,piani2008no}. 
 
 \par This study presents a simplified method for assessing zero quantum discord, which involves partitioning the density matrix into block matrices and evaluating the specific properties of these blocks.
Mathematically, the density operator $\rho_G$ is defined as \cite{hassan2007combinatorial,braunstein2006laplacian,joshi2018concurrence,joshi2022entanglement}: $$\rho_G = \frac{1}{Tr(L(G))} [L(G)]$$
where $L(G)$ represents the Laplacian matrix of the weighted graph $G$.

Since the density operator $\rho_G$ serves as an alternate representation of the state of a quantum system, it is pivotal in understanding the evolution of quantum systems, which are inherently unitary operations. Consequently, unitary operators, often called quantum gates, play a fundamental role in quantum physics. These gates, including single-qubit gates like the Pauli operators ($\sigma_x$, $\sigma_y$, $\sigma_z$), as well as multi-qubit gates, are crucial for manipulating and transforming quantum states \cite{nielsen2010quantum}. Notably, unitary operators enable the exploration of both separability and quantum discord, which are measures of the correlation between two subsystems of a quantum system. Ensuring the validity and reliability of quantum computations requires necessary conditions for a positive semi-definite matrix. These conditions encompass various mathematical criteria that must be satisfied by the matrix elements \cite{horn2012matrix}.

\textbf{Condition for positive semi-definite matrix:}
	Let $A= [a_{pq}]_{n \times n}$ be a positive semi-definite matrix then\begin{enumerate}
		\item  $|a_{ii}|  \geq \sum_{i \neq j} |a_{ij}|$ \cite{joshi2022entanglement}.
		\item   $ det( \begin{bmatrix}
			a_{pp} &a_{pq}\\ a_{qp}& a_{qq}
		\end{bmatrix}) \geq 0 $ for $q>p $ \cite{joshi2022entanglement}.
	\end{enumerate}  
	Converse is true if 	 $   \begin{bmatrix}
		\bar{x_p} & 	\bar{x_q}
	\end{bmatrix} \begin{bmatrix}
		a_{pp} &a_{pq}\\ a_{qp}& a_{qq}
	\end{bmatrix} \begin{bmatrix}
		{x_p} \\	{x_q}
	\end{bmatrix} = |b_{pq}| |x_p + (-1)^{m_{pq}} x_q|^2 + |c_{pq}| |x_p + (-1)^{n_{pq}} ix_q|^2 +[a_{pp}- |b_{pq}|- |c_{pq}|] |x_p|^2	+ [a_{qq}- |b_{pq}|- |c_{pq}|] |x_q|^2 \geq 0 $ for $q>p $, here $m_{pq}$ and $n_{pq}$ are either odd or even and  $|b_{pq}| $ and $|c_{pq}| $ are positive,
 
	and	  $   \begin{bmatrix}
		\bar{x_p} & 	\bar{x_r}
	\end{bmatrix} \begin{bmatrix}
		a_{pp} &a_{pr}\\ a_{rp}& a_{rr}
	\end{bmatrix} \begin{bmatrix}
		{x_p} \\	{x_r}
	\end{bmatrix} = |b_{pr}| |x_p + (-1)^{m_{pr}} x_r|^2 + |c_{pr}| |x_p + (-1)^{n_{pr}} ix_r|^2 +[a_{pp}- |b_{pr}|- |c_{pr}|] |x_p|^2	+ +[a_{rr}- |b_{pr}|- |c_{pr}|] |x_r|^2	 \geq 0 $ for $ r>p $, here $m_{pr}$ and $n_{pr}$ are either odd or even and  $|b_{pr}| $ and $|c_{pr}| $ are positive.
 
	Then there exists  $   \begin{bmatrix}
		\bar{x_q} & 	\bar{x_r}
	\end{bmatrix} \begin{bmatrix}
		a_{qq} &a_{qr}\\ a_{rq}& a_{rr}
	\end{bmatrix} \begin{bmatrix}
		{x_q} \\	{x_r}
	\end{bmatrix} = b_1 |x_q + (-1)^{m_{qr}} x_r|^2 + b_2 |x_q + (-1)^{m_{pq}+n_{pr}} \bar{i}x_r|^2 + b_3 |x_q + (-1)^{n_{pq}+m_{pr}} \bar{i}x_r|^2 + b_4 |x_q + (-1)^{n_{pq}+n_{pr}} i\bar{i} x_r|^2 + (+ve) $, where $b_1$, $b_2$, $b_3$, and $b_4$ $\geq 0$ and $m_{qr}$ is either odd or even and $m_{qr}= m_{pq}+  m_{pr}$, such that 
	$\sum  \begin{bmatrix}	\bar{x_p} & 	\bar{x_q}
	\end{bmatrix} \begin{bmatrix}
		a_{pp} &a_{pq}\\ a_{qp}& a_{qq}
	\end{bmatrix} \begin{bmatrix}
		{x_p} \\	{x_q}
	\end{bmatrix} = a_1|x_1 \pm x_2 \pm \dots \pm x_n|^2 + a_2 |x_1 \pm ix_2 \pm \dots \pm ix_n|^2 + (+ve)$ which clearly shows that the matrix $A $ is a positive semi definite matrix.\\

	\textbf{Quantum Gates:} Quantum gates display reversibility, ensuring that each input has a unique corresponding output within the reversible circuit. In the realm of quantum computing, a variety of gates exists, including single-qubit gates (SQG) such as $I_2$, $\sigma_x, \sigma_y, \sigma_z,$ and $H=\frac{1}{\sqrt{2}} [\sigma_x+\sigma_z]$, as well as multiqubit gates (MQG) such as Controlled-NOT (CX=$\begin{bmatrix} I_2&0\ 0 & \sigma_x \end{bmatrix}$), Controlled-Z (CZ=$\begin{bmatrix} I_2&0\ 0 &
\sigma_z\end{bmatrix}$), and Swap gates. Combining these single and multiqubit gates effectively enables global quantum computation \cite{divincenzo1995two,
	gu2021fast}. The impact of their operations on qubits are briefly described below:
\begin{enumerate}
\item   \textbf{Performing quantum gates on the graph}: Consider a graph $G$ with vertices $V(G)=\left\{v_1, v_2, \dots,v_n\right\}$ and edges $E(G)=\left\{e_{11}, e_{12},\dots,e_{nn}\right\}$. The vertices can be represented as column vectors ($b_i$). Applying a quantum gate to a graph  $G$ signifies the action of applying the quantum gate to individual vertices or edges.
\item   \textbf{Performing quantum gate operations on a density operator:} The expression $\rho_G
	=\frac{1}{Tr(L)} \big[ l_{ij} b_i \otimes {b_j}^T \big]_{2^n \times 2^n}$ represents the density operator, where $Tr(L)$ is the trace of matrix $L$, $l_{ij}$ is an
	element of matrix $L$, and $b_i$ and $b_j$ are basis vectors, then $\rho_G$
	$\xrightarrow{\text{ QG}} { \rho}' = \frac{1}{Tr(L)} \big[ l_{ij} (QG\times b_i\otimes {(QG \times b_j)}^T)
	\big]_{2^n \times 2^n}$ 			
			\item  \textbf{Partial quantum gate on a graph $G$\cite{joshi2024separability}:  } Consider a graph $G$ with $2^n$ vertices. Let $V$ be the collection of vertices, defined as $V=\{ b_i={c_1}^i\otimes {c_2}^i \otimes \dots \otimes {c_n}^i \text{ for all $i= 1,2,\dots ,2^n$}\}$, where $c_i's$ are column vectors in ${\mathbb{C}}^2$. Let us consider a unitary operator $U= I_2 \otimes {U_2 \otimes \dots \otimes U_n}$, where $U_2, \dots, U_n $ can be either $I_2$ or $\sigma_x$. In this case, a partial quantum gate on the graph can be defined as $(\text{Partial}U)G= (Ub_i,Ub_j) \in E(G_1)$ for all $(b_i,b_j) \in E(G)$, where, $U_k$ is determined by the following conditions, \begin{enumerate}
	\item $U_k = I_2$ if ${c_k}^i= {c_k}^j$ .
	 \item $U_k = \sigma_x$ if ${c_k}^i \neq {c_k}^j$.
        \end{enumerate} 
      \item 	 \textbf{Partial quantum gate on a density operator \cite{joshi2024separability}:} Let $\rho_G= \frac{1}{\text{Tr}(L(G))} \left[ l_{ij} b_i \otimes {b_j}^T \right]_{2^n \times 2^n}$ represent the density operator of the graph $G$ with $2^{n=p+q}$ vertices. The vertex set $V$ is defined as $V=\{ b_i={c_1}^i\otimes {c_2}^i \otimes \dots \otimes {c_n}^i \text{ for all $i= 1,2,\dots ,2^n$}\}$, where ${c_i}$'s are column vectors in ${\mathbb{C}}^2$. Let us consider a unitary operator $U= \underbrace{ I_2 \otimes \dots \otimes I_2}_{q} \otimes {U_{q+1} \otimes \dots \otimes U_{q+p} }$ such that $U_{q+1} , \dots, U_{q+p} $ are either $I_2$ or $\sigma_x$ and $\rho_G$ represents as a block matrix of dimensions $2^q \times 2^q $. In this case, a partial quantum gate on the density operator can be expressed as  $(\text{Partial}U) \rho_G= \frac{1}{Tr(L(G))} \big[  l_{ij} Ub_i \otimes {(Ub_j)}^T \big]_{2^n \times 2^n}$,  where, $U_k$ is determined by the following conditions,  \begin{enumerate}
	\item $U_k = I_2$ if ${c_k}^i= {c_k}^j$ .
	\item $U_k = \sigma_x$ if ${c_k}^i \neq {c_k}^j$.
\end{enumerate}
\end{enumerate}
\textit{In this article, we will adhere to the established definitions as:}
\begin{enumerate}
    \item  
  Consider $(G,a)$ as a graph linked to the $n$-qubit quantum state, consisting of $2^{n}$ ($n=p+q$) vertices labeled as $(ij)$, where $i$ varies from 1 to $2^p$, and $j$ varies from 1 to $2^q$.

\item The expression $|x|$ represents the absolute value of x.
\item For a matrix $A$, $\overline{A}$ represents as a conjugate of $A$.
\end{enumerate}
 
	The article is structured as follows:
 Section \ref{2} introduces a series of theorems about pure quantum states linked to weighted graphs. Section \ref{3} delves into the concept of discord and its intricate definition within the quantum realm. It further analyzes quantum discord, unraveling its underlying characteristics and implications from a graph-theoretic perspective. In Section \ref{4}, we present a conclusive summary, highlighting the significance of our research and its potential contributions to the evolving landscape of quantum information science.
	
	\section{Conditions for a positive semi-definite matrix and the pure state associated with a weighted graph} \label{2}
	In this section, we establish a set of theorems that are presented for pure states associated with weighted graphs. \\
	
	\begin{theorem} \label{T2.1}

  Let $(G,a)$ be a connected weighted graph, which represents an $m$-qubit state. A state $(G,a)$ is considered to be a pure state if the absolute value of $l_{ij}$ is equal to $d_{ii}$ for all values of $i$ and $j$.
  
	\end{theorem}
 
	Proof: Consider $(G,a)$ as a connected weighted graph, which represents an $m$-qubit state  
      and $\rho_G=\frac{1}{Tr(L(G))} L(G)= \frac{1}{Tr(L(G))} \begin{bmatrix}
		d_{11}	& l_{12 } &.&.&.& l_{12^m}\\
		l_{21}  & d_{22 }&.&.&.& l_{22^m}\\
		.  & , &.&.&.& .\\
		.  & , &.&.&.& .\\
		.  & , &.&.&.& .\\
		l_{2^m1}  & l_{2^m2} &.&.&.& d_{2^m2^m}\\
	\end{bmatrix}$ be the density operator. If $\abs{l_{ij}} = d_{ii}$ $\forall i,j$, then we clearly have $Tr({\rho_G}^2)=1$. Hence proved.\\

	\textbf{Example:}
The graphs $G_1$ and $G_2$ shown below also exemplify the theorem mentioned earlier.
       
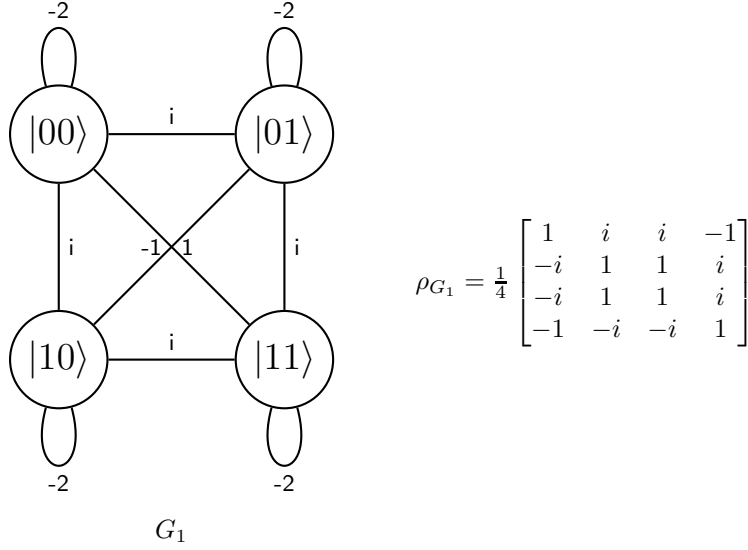
\begin{center}
           	\begin{tikzpicture}[auto, node distance=3cm, every loop/.style={},
		thick,main node/.style={circle,draw,font=\sffamily\Large\bfseries}]
		\node[main node] (1) {$\ket{00}$};
		\node[main node] (2) [ right of=1] {$\ket{01}$};
		\node[main node] (3) [ below of=1] {$\ket{10}$};
		\node[main node] (4) [ below of=2] {$\ket{11}$};
		\draw (1.5, -5) node[below] { $ \hspace{1cm} G_1 \hspace{1cm}$};   
  \draw (7, -1) node[below] { $ \hspace{1cm}\rho_{G_1} = \frac{1}{4} \begin{bmatrix}
    1 & i & i & -1 \\ -i & 1 & 1 & i \\ -i & 1 & 1 & i \\ -1 & -i & -i & 1
 \end{bmatrix} \hspace{1cm}$};   
		\path[every node/.style={font=\sffamily\small}]
		(1) edge node[] {i} (2)
                edge node[] {i} (3)
                edge node[  left] {-1} (4)
		edge [loop above ] node  {-2} (1)
          (2) edge node[ right ] {1} (3)
                edge node[] {i} (4)
               edge [loop above ] node  {-2} (2)
          (3) edge node[] {i} (4)
             edge [loop below ] node  {-2} (3)
		(4)edge [loop below ] node  {-2} (4);
	\end{tikzpicture}
       \end{center}

	Here $d_{11}= d_{22} =d_{33} = d_{44} = 1$, $l_{12}= l_{13}= l_{24}= l_{34}=i$, $l_{14}=l_{41}= -1$, $l_{21}= l_{31}= l_{42}= l_{43}=-i$ and $l_{23}=l_{32}=1$.
Clearly, we have $d_{ii}= \abs{l_{ij}}$. 

 
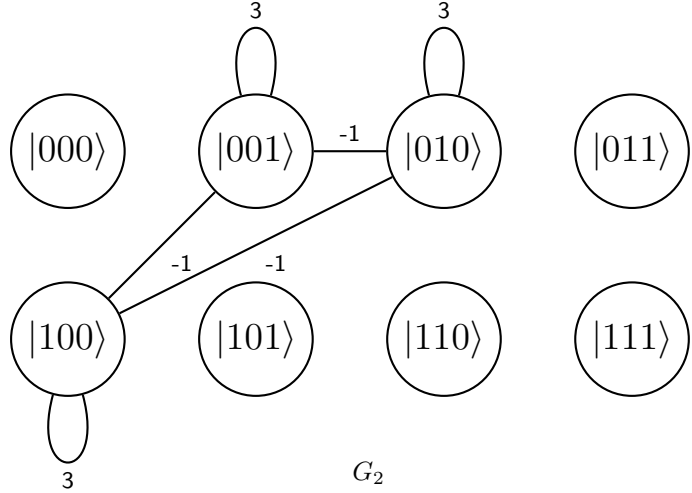
\begin{center}
			\begin{tikzpicture}[auto, node distance=2.5cm, every loop/.style={},
			thick,main node/.style={circle,draw,font=\sffamily\Large\bfseries}]
		    \node[main node] (1) {$\ket{000}$};
			\node[main node] (2) [ right of=1] {$\ket{001}$};
			\node[main node] (3) [ right of=2] {$\ket{010}$};
			\node[main node] (4) [ right  of=3] {$\ket{011}$};
			\node[main node] (5) [ below of=1] {$\ket{100}$};
			\node[main node] (6) [ below of=2] {$\ket{101}$};
			\node[main node] (7) [ below   of=3] {$\ket{110}$};
			\node[main node] (8) [ below   of=4] {$\ket{111}$};
			\draw (4,-4) node[below] { $G_2$ }; 
			\path[every node/.style={font=\sffamily\small}]
			(2) edge [] node[] {-1} (3)
			edge [] node[] {-1} (5)
			edge [loop above] node[] { 3} (1)
			(3) edge [] node[] {-1} (5)
			edge [loop above] node[] { 3} (3)
			(5) edge [loop below] node[] { 3} (5);
			\end{tikzpicture}
		\end{center}

 $$ \rho_{G_2} = \frac{1}{3} \begin{bmatrix}
     0 & 0 & 0 & 0 &  0 & 0 & 0 & 0\\ 0 & 1 & 1 & 0 &  1 & 0 & 0 & 0\\ 0 & 1 & 1 & 0 &  1 & 0 & 0 & 0 \\0 & 0 & 0 & 0 &  0 & 0 & 0 & 0\\ 0 & 1 & 1 & 0 &  1 & 0 & 0 & 0\\  0 & 0 & 0 & 0 &  0 & 0 & 0 & 0\\  0 & 0 & 0 & 0 &  0 & 0 & 0 & 0 \\  0 & 0 & 0 & 0 &  0 & 0 & 0 & 0
 \end{bmatrix}$$
 which further supports the above theorem.

	\begin{theorem} \label{T2.2}
		Consider a weighted graph $(G,a)$ with $2^m$ vertices, where $ a \in \C $. This graph is associated with an $m$-qubit state that has more than two components. If the graph consists of only one connected component, labeled as $G_1$, and each vertex in $G_1$ is linked to every other vertex via unique edges, satisfying the condition $\abs{a{(v_i,v_j)}}= a(v_i,v_i) +\sum_{v_i \neq v_j} \abs{a{(v_i,v_j)}}$ for all $v_i$, $v_j$ $\in G_1$, then, the weighted graph $(G,a)$ corresponds to a pure quantum state.
	\end{theorem}
	Proof: 	Let $(G,a)$  $ (a \in \C)$ be a weighted graph with $2^m$ vertices, which represents an $m$-qubit quantum state. Let $G_1$ be the connected component of $G$ that consists of $k$ vertices and absolute value of edge weight ($a{(v_i,v_j)}$, for $v_i \neq v_j $) is $c$. Suppose  $\abs{a{(v_i,v_j)}}= a(v_i,v_i) +\sum_{v_i \neq v_j} \abs{a{(v_i,v_j)}} $ for all $v_i$, $v_j$ $\in G_1$ which implies that diagonal sum of density operator is $kc$, and the diagonal entries of ${\rho_G}^2$ is $kc^2$. Therefore $Tr({\rho_G}^2)=1$.	Therefore, it has been demonstrated. \\
 
 \textbf{Example: } 	The graph $G$ depicted below serves as an illustration of theorem \ref{T2.2}.

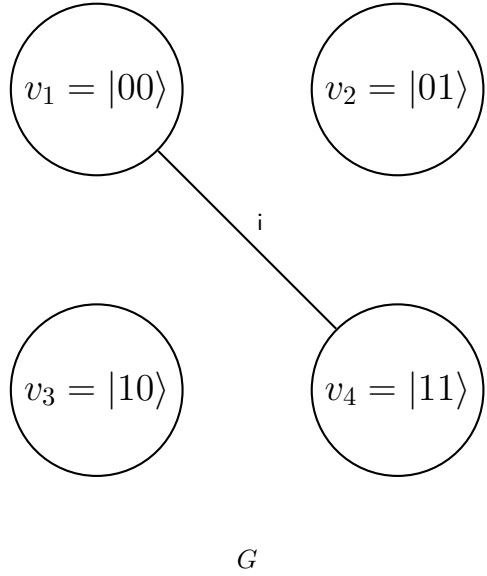
 \begin{center}
			\begin{tikzpicture}[auto, node distance=4cm, every loop/.style={},
			thick,main node/.style={circle,draw,font=\sffamily\Large\bfseries}]
		    \node[main node] (1) {$v_1=\ket{00}$};
			\node[main node] (2) [ right of=1] {$v_2=\ket{01}$};
			\node[main node] (3) [ below of=1] {$v_3=\ket{10}$};
			\node[main node] (4) [ below of=2] {$v_4=\ket{11}$};
			
			\draw (2,-6) node[below] { $G$ }; 
			\path[every node/.style={font=\sffamily\small}]
			(1) edge [] node[] {i} (4);
			\end{tikzpicture}
		\end{center}

 $$ \rho_{G} = \frac{1}{2} \begin{bmatrix}
    1 & 0 & 0 & i \\ 0 & 0 & 0 & 0 \\  0 & 0 & 0 & 0 \\ -i & 0 & 0 & 1
 \end{bmatrix}$$
Here connected component say $G_1$ is with vertex set $V=\{v_1,v_4\}$ and $a(v_1, v_1) = a(v_4, v_4) =0$, $a(v_1, v_4)=i$.

We can see that $\abs{a(v_1, v_4)} = a(v_1, v_1) + \sum_{v_i \neq v_j} \abs{a{(v_i,v_j)}}$

	\begin{corollary} \label{C2.2.1}
	
  Consider a connected weighted graph $(G,a)$ with $2^m$ vertices, associated with an $m$-qubit state . The weight of each edge is represented by a real number $a$. If the absolute of the weight of an edge from vertex $v_i$ to $v_j$ (where $v_i \neq v_j$) is equal to the total of the weights of all edges connected to vertex $v_i$, for all pairs of vertices $v_i$ and $v_j$, then the weighted graph $(G,a)$ represents the pure state.
	\end{corollary}

 \section{Quantum Discord of weighted Graphs} \label{3}
				
	In this section, we discuss various conditions for zero or nonzero quantum discord in the associated weighted graphs using block density operators.\\
			
	\begin{theorem} \label{T3.1}
  Consider the density operator $\rho_G = [\rho_{ij}]_{4 \times 4}$, which represents a weighted graph $(G,a)$ on ${\mathbb{C}}^2 \otimes {\mathbb{C}}^2$. Given that U is a unitary operator expressed as $U=I_2 \otimes U_2 $. If $ (\overline{\text{Partial}U)\rho_G}  =\rho_G$ and $ (\rho_{11}-\rho_{22} )\rho_{14}=\rho_{12} (\rho_{13}-\rho_{24} )$ then quantum discord of the weighted graph, $D_Q(G,a) =0$.

	\end{theorem}
 
Proof:  Let $(G,a)$ be a weighted graph defined on ${\mathbb{C}}^2 \otimes {\mathbb{C}}^2$ and $\rho_G=[\rho_{ij}]_{4 \times 4}$ represents the density operator. Clearly, $(\overline{\text{Partial}U) \rho_G} =\rho_G$  implies that blocks in the density operator are Hermitian matrices. Moreover, $(\rho_{11}-\rho_{22})\rho_{14}= \rho_{12}(\rho_{13}-\rho_{24})$ reveals commutativity of the blocks as expressed by $\rho_{11} \rho_{14}+ \rho_{12} \rho_{24}= \rho_{12}\rho_{13} +\rho_{22} \rho_{14}$. Hence the proof \cite{huang2011new}.\\
	
\textbf{Example:}
Let $\rho= \frac{1}{4}[ I + \sum_{i=1}^{3} T_{ii} \sigma_i \otimes \sigma_i ]$ be a density operator.The quantum discord for the density matrix $\rho$ is equal to zero if $T_{22}=T_{33}=0$ \cite{dakic2010necessary, yao2012geometric}.\\

 $
\rho= \frac{1}{4}[ I + \sum_{i=1}^{3} T_{ii}\sigma_i \otimes \sigma_i ] = \frac{1}{4} \begin{bmatrix}
1 + T_{33} & 0 & 0 & T_{11}-T_{22}\\
0 & 1-T_{33} &  T_{11}+T_{22} & 0 \\
0 & T_{11}+T_{22}&  1-T_{33} & 0\\
T_{11}-T_{22} & 0  & 0 & 1+T_{33}
\end{bmatrix}$
       \\
As per the Theorem \ref{T3.1} $(\overline{\text{Partial}U) \rho_G} =\rho_G$ if  $T_{11}-T_{22}=T_{11}+T_{22} $ which implies $T_{22}=0$. Also for $\rho_{11} \rho_{14}+ \rho_{12} \rho_{24}= \rho_{12}\rho_{13} +\rho_{22} \rho_{14}$, one can show  $(1+T_{33})(T_{11}-T_{22})=(1-T_{33})(T_{11}-T_{22})$. For $T_{22}=0$, $T_{11}+T_{33}T_{11}= T_{11}-T_{33}T_{11}$ or $2T_{33}(T_{11}) =0 $ which implies that $T_{33}=0$.\\

\begin{corollary} \label{C3.1.1}
 
  Let $\rho_G$ represent the density operator of a weighted graph $(G,a)$ on ${\mathbb{C}}^2 \otimes {\mathbb{C}}^2$. The quantum discord $D_Q(G,a) =0$ if $(\overline{I_2 \otimes \sigma_x){ \rho_G }}=\rho_G $ and  $(\sigma_x \otimes I_2) \rho_G =\rho_G $.

\end{corollary}
				
Proof:	
 Let $\rho_G$ represent the density operator  of a weighted graph $(G,a)$ on ${\mathbb{C}}^2 \otimes {\mathbb{C}}^2$, given by $\rho_G = \begin{bmatrix} A & B \\ B^* & C \end{bmatrix}$.
The  conditions $(\overline{I_2 \otimes \sigma_x){\rho_G}} = \rho_G$ and $(\sigma_x \otimes I_2) \rho_G = \rho_G$ show that each block is in the form $\begin{bmatrix} a & ib \\ -ib & a\end{bmatrix}$ and $\rho_{11}-\rho_{22} = \rho_{13}-\rho_{24} = 0$. Using these results, one can easily show that blocks are commuting with each other.   \cite{huang2011new}.

\begin{theorem} \label{T3.2} 
Let $\rho_G$ represent the density operator corresponding to a weighted graph $(G,a)$. Suppose the graph is associated with an $n$-qubit state ($n=p+q$) such that $\rho_G= [A^{xy}]_{2^q \times 2^q}$ $(A^{xy}= [a^{xy}_{ij}]_{2^p \times 2^p} )$. The density operator $\rho_G$ can recursively  represented as the block matrices such as
$A^{xy} = B^{xy} - C^{xy} + iD^{xy} - iE^{xy}$ where $B^{xy}$, $C^{xy}$, $D^{xy}$, and $E^{xy}$ $\geq 0$.
The quantum discord $D_Q(G,a)=0$ if $B^{lm}$, $B^{rs}$, $C^{lm}$, $C^{rs}$,$D^{lm}$, $D^{rs}$, $E^{lm}$, and $E^{rs}$ all commute with each other, for all $lm$ and $rs$ where $A^{rs} = B^{rs} - C^{rs} + iD^{rs} - iE^{rs}$ with $B^{rs}$, $C^{rs}$, $D^{rs}$, and $E^{rs}$ $\geq 0$, and $A^{lm} = B^{lm} - C^{lm} + iD^{lm} - iE^{lm}$ where $B^{lm}$, $C^{lm}$, $D^{lm}$, and $E^{lm}$ $\geq 0$.
\end{theorem}

Proof: Consider $\rho_G$ as the density operator corresponding to a weighted graph $(G,a)$. Let us assume the graph is associated with an $n$-qubit state, and $\rho_G$ can be represented as a block matrix $\rho_G = [A^{xy}]{2^q \times 2^q}$, where $A^{xy} = [a^{xy}_{ij}]_{2^p \times 2^p}$. Assume that $A^{lm} = B^{lm} - C^{lm} + iD^{lm} - iE^{lm}$, where $ B^{lm}$,$C^{lm}$,$D^{lm}$, and $E^{lm}$ are positive semi-definite matrices and $A^{rs} = B^{rs} - C^{rs} + iD^{rs} - iE^{rs}$ where $B^{rs}$, $C^{rs}$, $D^{rs}$, and $E^{rs}$ are positive semi-definite matrices. If $B^{lm}$, $B^{rs}$, $C^{lm}$, $C^{rs}$,$D^{lm}$, $D^{rs}$, $E^{lm}$, and $E^{rs}$ all commute with each other, for all $lm$ and $rs$, then $A^{lm}A^{rs}= (B^{lm} - C^{lm} + iD^{lm} - iE^{lm})(B^{rs} - C^{rs} + iD^{rs} - iE^{rs}) = (B^{lm} - C^{lm} + iD^{lm} - iE^{lm})(B^{rs}) -(B^{lm} - C^{lm} + iD^{lm} - iE^{lm})(C^{rs}) +i (B^{lm} - C^{lm} + iD^{lm} - iE^{lm})(D^{rs} )- i (B^{lm} - C^{lm} + iD^{lm} - iE^{lm})(E^{rs})= (B^{rs})(B^{lm} - C^{lm} + iD^{lm} - iE^{lm}) -(C^{rs})(B^{lm} - C^{lm} + iD^{lm} - iE^{lm}) +i (D^{rs})(B^{lm} - C^{lm} + iD^{lm} - iE^{lm})- i(E^{rs}) (B^{lm} - C^{lm} + iD^{lm} - iE^{lm})= (B^{rs} - C^{rs} + iD^{rs} - iE^{rs}) (B^{lm} - C^{lm} + iD^{lm} - iE^{lm}) =A^{rs}A^{lm}$. Hence Proved \cite{huang2011new}.\\

\textbf{Example:} The graph $G$ shown below illustrates the aforementioned theorem.\\
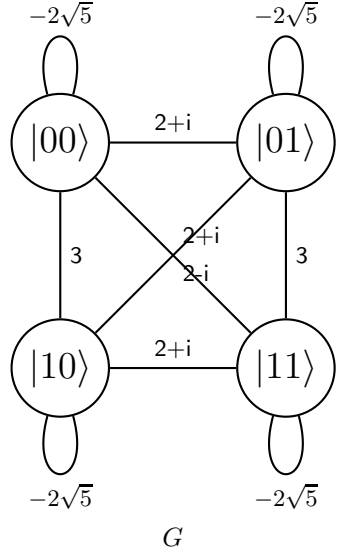
	$$	\begin{tikzpicture}[auto, node distance=3cm, every loop/.style={},
		thick,main node/.style={circle,draw,font=\sffamily\Large\bfseries}]
		\node[main node] (1) {$\ket{00}$};
		\node[main node] (2) [ right of=1] {$\ket{01}$};
		\node[main node] (3) [ below of=1] {$\ket{10}$};
		\node[main node] (4) [ below of=2] {$\ket{11}$};
      \draw (1.5, -5) node[below] { $ \hspace{1cm} G \hspace{1cm}$};   	    	 
	\path[every node/.style={font=\sffamily\small}]
    (1) edge node  [] {2+i} (2)
	(1) edge node  [] {3} (3)
	(1) edge node  [] {2+i} (4)
	(1) edge [loop above] node[]{ $-2\sqrt{5}$} (1)
	(2) edge node  [] {2-i} (3)
	(2) edge node  [] {3} (4)	
	(2) edge [loop above] node[]{$-2\sqrt{5}$} (2)
	(3)edge node  [] {2+i} (4) 
	(3) edge [loop below] node[]{ $-2\sqrt{5}$} (3)
	(4) edge [loop below] node[]{ $-2\sqrt{5}$} (4);
	\end{tikzpicture} $$\\
													
	We have	$\rho_G = \frac{1}{12} \begin{bmatrix}
		3 & 2+i  & 3 & 2+i \\ 2-i & 3 & 2-i & 3 \\ 3 & 2+i & 3 & 2+i  \\ 2-i  & 3 & 2-i & 3
		\end{bmatrix} = \begin{bmatrix}
		A^{11} & A^{12}\\ A^{21} & A^{22}
	\end{bmatrix} =	\frac{1}{3}\Big\{ \frac{1}{2}  \begin{bmatrix}1 & 1\\1 & 1	\end{bmatrix} \otimes  \frac{1}{2}\begin{bmatrix}
	1&i\\-i&1	\end{bmatrix} \Big\} +\frac{2}{3}\Big\{  \frac{1}{4}  \begin{bmatrix}
		2 & 2\\2 & 2
\end{bmatrix} \otimes \frac{1}{2}  \begin{bmatrix}
1 & 1\\1 & 1
\end{bmatrix}\Big\}$. \\
Here $ (\rho_{11}-\rho_{22} )\rho_{14}=\rho_{12} (\rho_{13}-\rho_{24} ) $, 
 $A^{11}=  \frac{1}{12}\begin{bmatrix}
3 & 2+i  \\ 2-i & 3
\end{bmatrix} $, $A^{12}= \frac{1}{12}\begin{bmatrix}
3 & 2+i  \\ 2-i & 3
\end{bmatrix}= \frac{1}{12} \Bigg\{ \begin{bmatrix} 2&2\\2&2\end{bmatrix}  
+ \begin{bmatrix}
1 & i\\-i & 1
\end{bmatrix} \Bigg \}
$, \\
and $A^{lm} A^{rs}=A^{rs}A^{lm}$.\\

\begin{theorem} \label{T3.3} 

Consider $\rho_G$ as the density operator of a weighted graph $(G,a)$ that is associated with an $n$-qubit state such that $\rho_G= [A^{xy}]_{2^q \times 2^q}$ $(A^{xy}= [a^{xy}_{ij}]_{2^p \times 2^p} )$ can be represented as a block matrix. Assume $(\overline{PartialU) {\rho_G} }=\rho_G$,  where $U=\underbrace{I_2 \otimes \dots \otimes I_2}_{q} \otimes \underbrace{ U_{q+1} \dots \otimes  U_{q+p} }_{p} $, and $U_i$ represents either $I_2$ or $\sigma_x$.
The quantum discord $D_Q(G,a)=0$, if $  \overline{(PartialU') {A^{rs} A^{lm}}}= A^{rs} A^{lm}$ for all $lm$ and $rs$, where $U' =(U_{1} \otimes U_{2} \otimes \dots \otimes U_{p})$.

\end{theorem}
Proof: 
Consider $\rho_G$  as the density operator of a weighted graph $(G,a)$ which is associated with an $n$-qubit state.
Assume that the density operator $\rho_G$ can be expressed as a block matrix, denoted as $[A^{xy}]_{2^q \times 2^q}$, where each block $A^{xy}$ is a $2^p \times 2^p$ matrix represented by $[a^{xy}_{ij}]_{2^p \times 2^p}$. Consider a unitary operator $U$ defined as:  $U=\underbrace{I_2 \otimes \dots \otimes I_2}_{q} \otimes \underbrace{ U_{q+1} \dots \otimes  U_{q+p} }_{p} $, where $U_i$ are either $I_2$ or $ \sigma_x  $.
Furthermore, if $(\overline{PartialU) {\rho_G} }=\rho_G$, it implies that the blocks  $A^{xy}$ being Hermitian.
Now if $ \overline{(PartialU') {A^{rs} A^{lm}}}= A^{rs} A^{lm}$ for all $lm$ and $rs$, where $U' =(U_{1} \otimes U_{2} \otimes \dots \otimes U_{p})$ then $A^{rs} A^{lm}$ is Hermitian. Consequently, it also implies that $A^{rs} A^{lm}= A^{lm}A^{rs}$. Hence $D_Q(G,a)=0$ \cite{huang2011new}.

\textbf{Example:} 
The graphs shown below, denoted as $G_1$ and $G_2$, also exemplify the theorem stated above.
$$
{	\begin{tikzpicture}[auto, node distance=4cm, every loop/.style={},
		thick,main node/.style={circle,draw,font=\sffamily\Large\bfseries}]
		\node[main node] (1) {$\ket{00}$};
		\node[main node] (2) [ right of=1] {$\ket{01}$};
		\node[main node] (3) [ below of=1] {$\ket{10}$};
		\node[main node] (4) [ below of=2] {$\ket{11}$};
      \draw (2.5, -5) node[below] { $ \hspace{1cm} G_1 \hspace{1cm}$};   	    	 
	\path[every node/.style={font=\sffamily\small}]
   	(1) edge node  [] {1} (3)
	(1) edge node  [above left] {-i} (4)
	(2) edge node  [above right] {i} (3)
	(2) edge node  [] {1} (4);
	\end{tikzpicture}}$$
The density operator $\rho_G$ for the graph $G_{1}$ is,

$$ \rho_{G_1} = \frac{1}{8}\begin{bmatrix} 2 & 0 & 1 & -i \\ 0 & 2 & i & 1 \\ 1 & -i & 2 & 0 \\ i & 1 & 0 & 2 \end{bmatrix} $$

 Here $A^{11}=  \frac{1}{8} \begin{bmatrix} 2 & 0 \\ 0 & 2 \end{bmatrix} $ and $ A^{12}=\frac{1}{8}\begin{bmatrix} 1 & -i \\ i & 1 \end{bmatrix} $. \\
 Therefore, we can show that $ \overline{(PartialU') {A^{11} A^{12}}} =  A^{11} A^{12} = \frac{1}{32}\begin{bmatrix}  1 & -i \\ i & 1 \end{bmatrix} = A^{12} A^{11}$.

Moreover, for the graph $G_{2}$
 $$	
{\begin{tikzpicture}[auto, node distance=3cm, every loop/.style={},
		thick,main node/.style={circle,draw,font=\sffamily\Large\bfseries}]
		\node[main node] (1) {$\ket{000}$};
		\node[main node] (2) [ right of=1] {$\ket{001}$};
		\node[main node] (3) [ right of=2] {$\ket{010}$};
		\node[main node] (4) [ right of=3] {$\ket{011}$};
            \node[main node] (5) [ below of=1] {$\ket{100}$};
		\node[main node] (6) [below of= 2] {$\ket{101}$};
		\node[main node] (7) [below of= 3] {$\ket{110}$};
		\node[main node] (8) [below of= 4] {$\ket{111}$};
      \draw (4.5, -5) node[below] { $ \hspace{1cm} G_2 \hspace{1cm}$};   	    	 \path[every node/.style={font=\sffamily\small}]
   	(1) edge node  [] {} (5)
	(1) edge node  [] {} (7)
	(1) edge node  [] {} (8)
	(2) edge node  [] {} (6)
	(2) edge node  [] {} (7)
	(2) edge node  [] {} (8)
        (3) edge node  [] {} (5)
	(3) edge node  [] {} (6)
	(3) edge node  [] {} (7)
	(4) edge node  [] {} (5)
	(4) edge node  [] {} (6)
	(4) edge node  [] {} (8);
	\end{tikzpicture}} $$

{the density operator $\rho_{G_2}$ is,
$$ \rho_{G_2} = \frac{1}{24}\begin{bmatrix} 3 & 0 & 0 & 0 & -1 & 0 & -1 & -1\\ 0 & 3 & 0 & 0 & 0 & -1 & -1 & -1\\ 0 & 0 & 3 & 0 & -1 & -1 & -1 & 0 \\ 0 & 0 & 0 & 3 & -1 & -1 & 0 & -1\\  -1 & 0 & -1 & -1 & 3 & 0 & 0 & 0 \\ 0 & -1 & -1 & -1 & 0 & 3 & 0 & 0 \\  -1 & -1 & -1 & 0 & 0 & 0 & 3 & 0 \\ -1 & -1 & 0 & -1 & 0 & 0 & 0 & 3 \end{bmatrix} $$
where $A^{11}=  \frac{1}{24} \begin{bmatrix} 3 & 0 & 0 & 0 \\ 0 & 3 & 0 & 0 \\ 0 & 0 & 3 & 0 \\ 0 & 0 & 0 & 3   \end{bmatrix} $ and $ A^{12}=\frac{1}{24}\begin{bmatrix} -1 & 0 & -1 & -1\\ 0 & -1 & -1 & -1 \\ -1 & -1 & -1 & 0 \\ -1 & -1 & 0 & -1 \end{bmatrix} $,\\
Similar to the case of $G_{1}$, we can show that $ \overline{(PartialU') {A^{11} A^{12}}} =  A^{11} A^{12} = \frac{1}{576}\begin{bmatrix} -3 & 0 & -3 & -3 \\ 0 & -3 & -3 & -3 \\ -3 & -3 & -3 & 0 \\ - 3 & -3 & 0 & -3 \end{bmatrix} =A^{12} A^{11}$.}
 

\begin{theorem}  \label{T3.4}
Consider $\rho_G$ as the density operator of a weighted graph $(G,a)$ connected to an $n$-qubit state with $2^{n}$ vertices, where $\rho_G= [A^{xy}]_{2^{n-1} \times 2^{n-1}}$ $(A^{xy}= [a^{xy}_{ij}]_{2 \times 2} )$. If $\rho_G = \overline{(\text{Partial}U) \rho_G}$  with $\abs{a^{xy}_{ij}}= \abs{a^{xy}_{ij+1}}$ where $U=\underbrace{ I_2 \otimes \dots \otimes I_2}_{n-1}\otimes U_n$ is a unitary operator, then the graph corresponding to the quantum state will exhibit a quantum discord value of zero.


\end{theorem} 

Proof: Consider $\rho_G$ as the density operator of a weighted graph $(G,a)$ connected to an $n$-qubit state with $2^{n}$ vertices
where $\rho_G= [A^{xy}]_{2^{n-1} \times 2^{n-1}}$ $(A^{xy}= [a^{xy}_{ij}]_{2 \times 2} )$.
Let us further introduce a unitary operators $U=\underbrace{ I_2 \otimes \dots \otimes I_2}_{n-1}\otimes U_n$, where ${U_{n}}$ is either $I_2$ or $\sigma_x$. 
For $\rho_G = \overline{(\text{Partial}U) \rho_G}$, we have all the $2$-dimensional blocks $A^{xy}$ of $\rho_G$ as Hermitian matrices. Furthermore, if   $\abs{a^{xy}_{ij}}= \abs{a^{xy}_{ij+1}}$ then  block $A^{xy}$ either takes the form $A^{xy}= i^m\begin{bmatrix}
a&ia\\-ia&a
\end{bmatrix}$  
or $A^{xy}=\begin{bmatrix}
a&a\\a&a
\end{bmatrix}$.
Hence, the proof \cite{huang2011new}.\\
\textbf{Example:} The graphs illustrated below, denoted as $G_1$ and $G_2$, also exemplify the theorem discussed above.
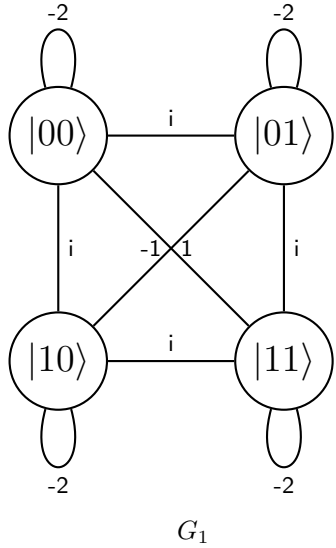
\begin{center} 
           	\begin{tikzpicture}[auto, node distance=3cm, every loop/.style={},
		thick,main node/.style={circle,draw,font=\sffamily\Large\bfseries}]
		\node[main node] (1) {$\ket{00}$};
		\node[main node] (2) [ right of=1] {$\ket{01}$};
		\node[main node] (3) [ below of=1] {$\ket{10}$};
		\node[main node] (4) [ below of=2] {$\ket{11}$};
		\draw (1.8, -5) node[below] { $ \hspace{1cm} G_1 \hspace{1cm}$};   	    	 
		\path[every node/.style={font=\sffamily\small}]
		(1) edge node[] {i} (2)
                edge node[] {i} (3)
                edge node[  left] {-1} (4)
		edge [loop above ] node  {-2} (1)
          (2) edge node[ right ] {1} (3)
                edge node[] {i} (4)
               edge [loop above ] node  {-2} (2)
          (3) edge node[] {i} (4)
             edge [loop below ] node  {-2} (3)
		(4)edge [loop below ] node  {-2} (4);
	\end{tikzpicture}
       \end{center}
       The density operator linked to the graph $G_{1}$ can be expressed as
$${ \rho_{G_1} = \frac{1}{4} \begin{bmatrix}
    1 & i & i & -1 \\ -i & 1 & 1 & i \\ -i & 1 & 1 & i \\ -1 & -i & -i & 1
 \end{bmatrix} }$$
 Thus, $A^{11} = \frac{1}{4} \begin{bmatrix} 1 & i \\ -i & 1 \end{bmatrix}$, $A^{12}= \frac{1}{4} i\begin{bmatrix} 1 & i \\ -i & 1 \end{bmatrix}$ which shows $A^{11}A^{12}= A^{12}A^{11}.$

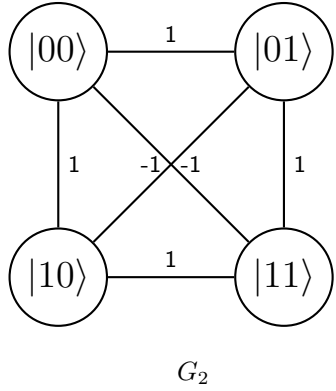
\begin{center}
 	\begin{tikzpicture}[auto, node distance=3cm, every loop/.style={},
		thick,main node/.style={circle,draw,font=\sffamily\Large\bfseries}]
		\node[main node] (1) {$\ket{00}$};
		\node[main node] (2) [ right of=1] {$\ket{01}$};
		\node[main node] (3) [ below of=1] {$\ket{10}$};
		\node[main node] (4) [ below of=2] {$\ket{11}$};
		\draw (1.8, -4) node[below] { $ \hspace{1cm} G_2 \hspace{1cm}$};   	    	 
		\path[every node/.style={font=\sffamily\small}]
		(1) edge node[] {1} (2)
                edge node[] {1} (3)
                edge node[  left] {-1} (4)
		  (2) edge node[ right ] {-1} (3)
                edge node[] {1} (4)
            (3) edge node[] {1} (4);
	\end{tikzpicture} 
       \end{center} 
 We can deduce the density operator associated with the graph $G_{2}$ as  
$${\rho_{G_2} = \frac{1}{4} \begin{bmatrix}
    1 & -1 & -1 & 1 \\ -1 & 1 & 1 & -1 \\ -1 & 1 & 1 & -1 \\ 1 & -1 & -1 & 1
 \end{bmatrix}}$$
Evidently, $A^{11} = \frac{1}{4} \begin{bmatrix} 1 & -1 \\ -1 & 1 \end{bmatrix}$, $A^{12}= -\frac{1}{4} \begin{bmatrix} 1 & -1 \\ -1 & 1 \end{bmatrix}$ indicating $A^{11}A^{12}= A^{12}A^{11}$.\\

\begin{theorem} \label{T3.5}

Consider $(G,a)$ as a weighted graph with $2^n$ vertices, which represents an $n$-qubit quantum state.
Let $E(G)$ be the set of edges in $(G,a)$, defined as $$E(G)= \{(v_{ij},v_{kl}); \text{  such that edge lies between
} v_{ij} \text{ and } v_{kl} \}.$$ If there exists an edge $(v_{ij},v_{kl})$ in graph G such that both $j$ and $l$ are either odd or even, then $D_Q(G,a)$, is equal to zero.

\end{theorem}
Proof:  Consider $(G,a)$ as a weighted graph with $2^n$ vertices, which represents an $n$-qubit quantum state. Let $E(G)$ be the edge set in $(G,a)$, defined as $E(G)= \{(v_{ij},v_{kl}); \text{  such that edge lies between
} v_{ij} \text{ and } v_{kl} \}.$

Interestingly, if $(v_{ij},v_{kl})$ belongs to $E(G)$ and both $j$ and $l$ are either odd or even, then the density operator $(\rho_G)$ can be expressed as a diagonal matrix in every block.

Diagonal matrices have a significant property that they commute with each other. Therefore, the blocks of the density operator $(\rho_G)$, being diagonal matrices, also commute. Hence, it is proved \cite{huang2011new}.\\~\\

\textbf{Example:} The graph displayed below, represented by $G$, offers an illustration of the theorem mentioned earlier

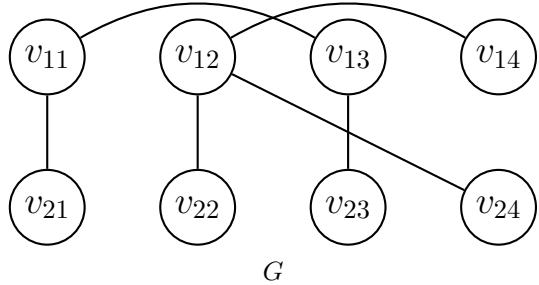
\begin{center}
         	\begin{tikzpicture}[auto, node distance=2cm, every loop/.style={},
		thick,main node/.style={circle,draw,font=\sffamily\Large\bfseries}]
		\node[main node] (1) {$v_{11}$};
		\node[main node] (2) [ right of=1] {$v_{12}$};
		\node[main node] (3) [ right of=2] {$v_{13}$};
		\node[main node] (4) [ right of=3] {$v_{14}$};
            \node[main node] (5) [ below of=1] {$v_{21}$};
		\node[main node] (6) [ below of=2] {$v_{22}$};
		\node[main node] (7) [ below of=3] {$v_{23}$};
		\node[main node] (8) [ below of=4] {$v_{24}$};
		\draw (3, -2.6) node[below] { $ \hspace{1cm} G\hspace{1cm}$};   	  \path[every node/.style={font=\sffamily\small}]
		(1) edge [bend left] node[] {} (3)
                edge node[] {} (5)
          (2) edge [bend left] node[] {} (4)
              edge node[] {} (6)
              edge node[] {} (8)
           (3) edge node[] {} (7);
	\end{tikzpicture} 
       \end{center} 
  The associated density operator is $\rho_G = \frac{1}{12} \begin{bmatrix} 2 & 0 & -1 & 0 & -1 & 0 & 0 & 0 \\ 0 & 3 & 0 & -1 & 0 & -1 & 0 & -1 \\ -1 & 0 & 2 & 0 & 0 & 0 & -1 & 0 \\ 0 & -1 & 0 & 1 & 0 & 0 & 0 & 0 \\ -1 & 0 & 0 & 0 & 1 & 0 & 0 & 0 \\ 0 & -1 & 0 & 0 & 0 & 1 & 0 & 0 \\ 0 & 0 & -1 & 0 & 0 & 0 & 1 & 0 \\ 0 & -1 & 0 & 0 & 0 & 0 & 0 & 1 \end{bmatrix}$
where each $2 \times 2$ block are in the diagonal form.\\

\begin{theorem} \label{T3.6}

Consider a weighted graph $(G,a)$ with $2^n$ vertices, associated with an $n$-qubit quantum state. Let $E(G)$ denote the set of edges in $(G,a)$, defined as  $$E(G)= \{(v_{ij},v_{kl}) \text{  such that edge lies between} v_{ij} \text{ and } v_{kl} \}.$$ If there exists an edge $(v_{ij},v_{kl})$ in the graph $(G,a)$ for all $i=k$, then the graph $(G,a)$ is correlated with a quantum state that has zero discord.

\end{theorem}
Proof:
Consider a weighted graph $(G, a)$ having $2^n$ vertices. This graph depicts a quantum state consisting of $n$-qubits. The set of edges of the graph $(G,a)$, represented as $$E(G)= \{(v_{ij},v_{kl}) \text{  such that edge lies between
} v_{ij} \text{ and } v_{kl} \}.$$ If $i=k$, then the density operator expressed as a diagonal block matrix which linked to the graph $(G,a)$. Notably, these diagonal matrices in the density operator are positive semi-definite.  Hence Proved.  \cite{huang2011new}.\\

\textbf{Example:} Below is a graph, labeled as $G$, which serves as an illustrative example of the previously discussed theorem.
$$	 
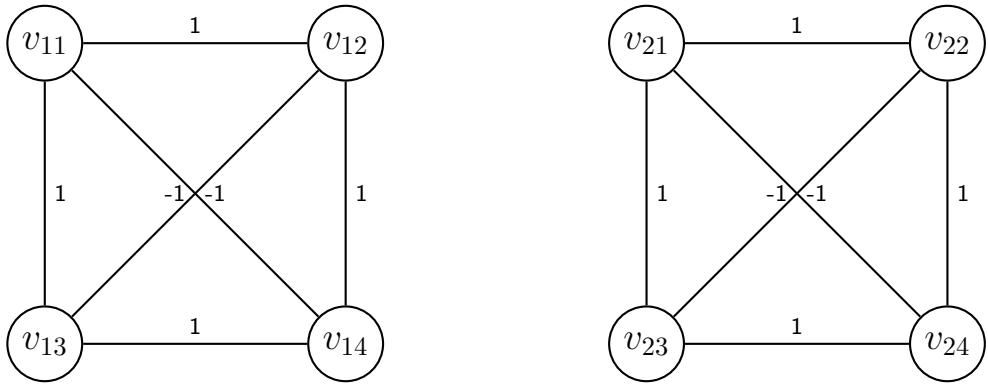
\begin{tikzpicture}[auto, node distance=4.0cm, every loop/.style={},
		thick,main node/.style={circle,draw,font=\sffamily\Large\bfseries}]
		\node[main node] (1) {$v_{11}$};
		\node[main node] (2) [ right of=1] {$v_{12}$};
		\node[main node] (3) [ right of=2] {$v_{21}$};
		\node[main node] (4) [ right of=3] {$v_{22}$};
		\node[main node] (5) [ below of=1] {$v_{13}$};
		\node[main node] (6) [ below of=2] {$v_{14}$};
		\node[main node] (7) [ below of=3] {$v_{23}$};
		\node[main node] (8) [ below of=4] {$v_{24}$};
		\draw (6,-6) node[below] { $ \hspace{1cm} $G$ \hspace{1cm}$};
		\path[every node/.style={font=\sffamily\small}]
		(1) edge [] node[] {1} (2)
		edge [left] node[] {-1} (6)
		edge [] node[] {1} (5)
		(2) edge [right] node[] {-1} (5)
		edge [] node[] {1} (6)
		(3) edge [] node[] {1} (4)
		edge  node[] {1} (7)
		edge  node[ left] {-1} (8)
		(4) edge [right] node[] {-1} (7)
		edge  node[] {1} (8)
		(5) edge [] node[] {1} (6)
		(7) edge  node[] {1} (8);
	\end{tikzpicture}$$	    
	
The density operator of the graph is a diagonal block matrix such that 
\begin{center}
$$\rho_G = \frac{1}{8} \begin{bmatrix} 1 & -1 & -1 & 1 & 0 & 0 & 0 & 0 \\ 
		-1 & 1 & 1 & -1 & 0 & 0 & 0 & 0  \\  -1 & 1 & 1 & -1  & 0 & 0 & 0 & 0 \\
		1 & -1 & -1 & 1  & 0 & 0 & 0 & 0  \\ 0 & 0 & 0 & 0 & 1 & -1 & -1 & 1 \\ 
		0 & 0 & 0 & 0 & -1 & 1 & 1 & -1  \\ 0 & 0 & 0 & 0 & -1 & 1 & 1 & -1   \\ 
		0 & 0 & 0 & 0 & 1 & -1 & -1 & 1 \end{bmatrix} $$
  \end{center} 
The proof, therefore, follows from the theorem.

\begin{theorem} \label{T3.7}
Consider $\rho_G = [A^{xy}]_{2 \times 2}$ as the density operator of a weighted graph $(G,a)$ with $2^n$ vertices, related to an 
$n$-qubit quantum state. Here, $A^{xy}$ is a block matrix defined as $A^{xy}=[A_{ij}^{xy}]_{2^{n-q-1} \times 2^{n-q-1}}$.
 
The quantum discord $D_Q(G,a)=0$ if 
$\sum_{k } {A_{ik}}^{xy}{A_{kj}}^{rs}=  
\sum_{k }  {{A_{ik}}^{rs}{A_{kj}}^{xy}}$ for all $i,j$
and $\overline{Partial(U^i)\rho_G}= \rho_G$, for $U^1=\underbrace{I_2 \otimes \dots \otimes I_2}_{p} \otimes  U_{p+1} \dots \otimes  U_{p+q}  $, and   $U^2=I_2 \otimes U_2 \dots \otimes   U_{n}  $, where, 
$U_k$ are either $I_2$ or $\sigma_x$.
\end{theorem}
Proof: Consider a weighted graph $(G,a)$ with $2^n$ vertices related to an $n$-qubit quantum state. Let $\rho_G = [A^{xy}]{2 \times 2}$ represent the density operator of this graph, where $A^{xy}$ is a block matrix of size $ 2^{n-q-1} $ and $A^{xy}$ is composed of elements denoted as $A{ij}^{xy}$. If $\sum_{k } {A_{ik}}^{xy}{A_{kj}}^{rs}=  
\sum_{k }  {{A_{ik}}^{rs}{A_{kj}}^{xy}}$ for all $i,j$ and  $\overline{Partial(U^i)\rho_G}= \rho_G$, for $U^1=\underbrace{I_2 \otimes \dots \otimes I_2}_{p} \otimes  U_{p+1} \dots \otimes  U_{p+q}  $, and   $U^2=I_2 \otimes U_2 \dots \otimes   U_{n}  $, where, 
$U_i$ are either $I_2$ or $\sigma_x$, then $A^{xy}$ and $A^{mn}$ commute, and each block of $A^{xy}$  also commutes. Hence, the discord is zero \cite{huang2011new}. \\

\begin{corollary} \label{C3.7.1}
 Consider  $\rho_G = [A^{xy}]_{4 \times 4}$ as the density operator of a weighted graph  $(G,a)$ with $2^n$ vertices where $(G,a)$ represents an $n$-qubit quantum state, and $A^{xy}$ be the block matrix of the density operator. The quantum discord $D_Q(G,a)=0$ if $(A^{11}-A^{22} )A^{14}=  A^{12}( A^{13} - A^{24} )$.
\end{corollary}

\begin{theorem} \label{T3.8}
Consider $ \rho_G = [\rho_{ij}]_{2^n \times 2^n}$ as the density operator of a weighted graph $(G,a)$ with $2^n$ vertices where $(G,a)$ represents an $n$-qubit quantum state. The quantum discord is zero, if the density operator can be expressed as $ \rho_G= \sum_{i} p_i {\rho_i}^1 \otimes {\rho_i}^2 $, with the condition that ${\rho_i}^1$ and ${\rho_i}^2$ commute.

\end{theorem}
Proof: 
Consider $ \rho_G = [\rho_{ij}]_{2^n \times 2^n}$ as the density operator of a weighted graph $(G,a)$ with $2^n$ vertices where $(G,a)$ represents an $n$-qubit quantum state.
Therefore, if the density operator is written as $ \rho_G= \sum_{i} p_i {\rho_i}^1 \otimes {\rho_i}^2$ where  ${\rho_i}^1$ and ${\rho_i}^2$ commute, then the quantum discord is zero. (The normality and commutativity of blocks of the density operator imply that the quantum discord is zero \cite{huang2011new}.)

\begin{theorem} \label{T3.9}
 Let $(G,a)$ represent a weighted graph with $2^n$ vertices, which represents an $n$-qubit state. The state has zero quantum discord, if $a_{ij,kl} =b $ $\forall$ $(v_{ij},v_{kl})\in E(G)$  and $d_{ij,ij}= b( 2^n-1 )$ for all $v_{ij}$.

\end{theorem}
 
Proof: Let $(G,a)$ be a weighted graph on $2^n$ vertices, which represents an $n$-qubit state. For  $a_{ij,kl}=b$ $\forall$ $(v_{ij},v_{kl})\in E(G)$  and $d_{ij,ij}= b( 2^n-1 )$ for all $v_{ij}$, we have $a \in \mathbb{R}$ and $d_{ij,ij}=	\sum {a_{ij, kl}}$ which illustrates that the graph $(G,a)$ is a complete graph. Therefore, the density operator can be formulated as $\rho_G = \rho_1 \otimes \rho_2$ and all blocks commute. Thus demonstrated. \cite{huang2011new}.\\

\textbf{Example: } Here, we discuss graphs associated with a quantum state that has zero quantum discord, e.g.,   a complete graph, and a complete weighted graph $K(n, a)$, $a>0$ associated with a three-qubit state.

$$ 	\begin{tikzpicture}[scale=0.45]                
\vertex[label=](1) at (0,0)   {$v_{11}$};
\vertex[label=](2) at (8,0)   {$v_{12}$};
\vertex[label=](3) at (14,-6)  {$v_{13}$};
\vertex[label=](4) at (14,-12)  {$v_{14}$};
\vertex[label=](5) at (8,-18)  {$v_{21}$};
\vertex[label=](6) at (1,-18)  {$v_{22}$};
\vertex[label=](7) at (-6,-12)  {$v_{23}$};
\vertex[label=](8) at (-6,-6)  {$v_{24}$};
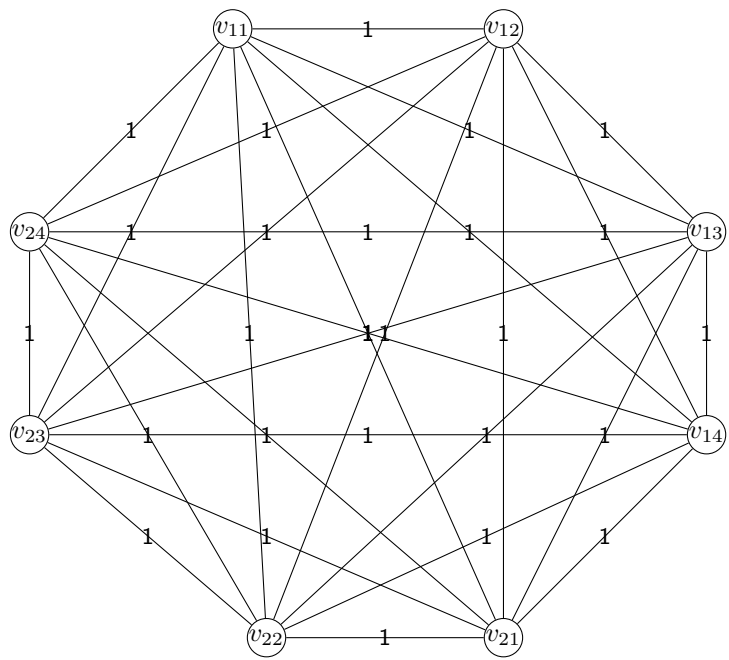
\draw (4,-22) node[below] { $ \hspace{1cm} \text{Complete Graph } (G_1) \hspace{1cm}$}; 	
\tikzset{EdgeStyle/.style={-}}
\path[every node/.style={font=\sffamily\small}]
(1) edge [] node[] {1} (2);
\path[every node/.style={font=\sffamily\small}]
(1) edge [] node[] {1} (3);
\path[every node/.style={font=\sffamily\small}]
(1) edge [] node[] {1} (4);
\path[every node/.style={font=\sffamily\small}]
(1) edge [] node[] {1} (5);
\path[every node/.style={font=\sffamily\small}]
(1) edge [] node[] {1} (6);
\path[every node/.style={font=\sffamily\small}]
(1) edge [] node[] {1} (7);
\path[every node/.style={font=\sffamily\small}]
(1) edge [] node[] {1} (8);
\path[every node/.style={font=\sffamily\small}]
(2) edge [] node[] {1} (3);
\path[every node/.style={font=\sffamily\small}]
(2) edge [] node[] {1} (4);
\path[every node/.style={font=\sffamily\small}]
(2) edge [] node[] {1} (5);
\path[every node/.style={font=\sffamily\small}]
(2) edge [] node[] {1} (6);
\path[every node/.style={font=\sffamily\small}]
(2) edge [] node[] {1} (7);
\path[every node/.style={font=\sffamily\small}]
(2) edge [] node[] {1} (8);
\path[every node/.style={font=\sffamily\small}]
(3) edge [] node[] {1} (4);
\path[every node/.style={font=\sffamily\small}]
(3) edge [] node[] {1} (5);
\path[every node/.style={font=\sffamily\small}]
(3) edge [] node[] {1} (6);
\path[every node/.style={font=\sffamily\small}]
(3) edge [] node[] {1} (7);
\path[every node/.style={font=\sffamily\small}]
(3) edge [] node[] {1} (8);
\path[every node/.style={font=\sffamily\small}]
(4) edge [] node[] {1} (5);
\path[every node/.style={font=\sffamily\small}]
(4) edge [] node[] {1} (6);
\path[every node/.style={font=\sffamily\small}]
(4) edge [] node[] {1} (7);
\path[every node/.style={font=\sffamily\small}]
(4) edge [] node[] {1} (8);
\path[every node/.style={font=\sffamily\small}]
(5) edge [] node[] {1} (6);
\path[every node/.style={font=\sffamily\small}]
(5) edge [] node[] {1} (7);
\path[every node/.style={font=\sffamily\small}]
(5) edge [] node[] {1} (8);
\path[every node/.style={font=\sffamily\small}]
(6) edge [] node[] {1} (7);
\path[every node/.style={font=\sffamily\small}]
(6) edge [] node[] {1} (8);
\path[every node/.style={font=\sffamily\small}]
(7) edge [] node[] {1} (8);
\end{tikzpicture}  $$

$$ 	\begin{tikzpicture}[scale=0.45]  
\vertex[label=](1) at (0,0)   {$v_{11}$};
\vertex[label=](2) at (8,0)   {$v_{12}$};
\vertex[label=](3) at (14,-6)  {$v_{13}$};
\vertex[label=](4) at (14,-12)  {$v_{14}$};
\vertex[label=](5) at (8,-18)  {$v_{21}$};
\vertex[label=](6) at (1,-18)  {$v_{22}$};
\vertex[label=](7) at (-6,-12)  {$v_{23}$};
\vertex[label=](8) at (-6,-6)  {$v_{24}$};
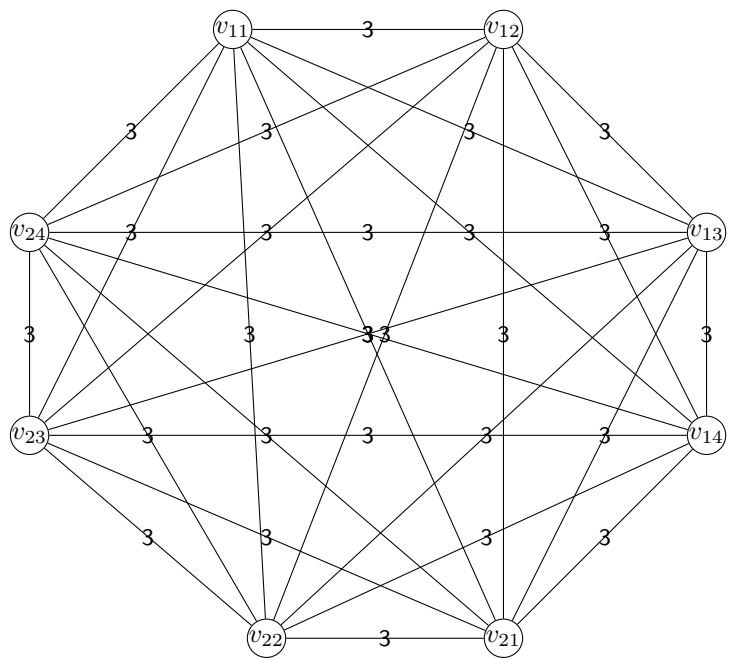
\draw (4,-21) node[below] { $ \hspace{1cm} \text{Complete Graph with weight $a=3$ } (G_2) \hspace{1cm}$}; 	
\tikzset{EdgeStyle/.style={-}}
\path[every node/.style={font=\sffamily\small}]
(1) edge [] node[] {3} (2);
\path[every node/.style={font=\sffamily\small}]
(1) edge [] node[] {3} (3);
\path[every node/.style={font=\sffamily\small}]
(1) edge [] node[] {3} (4);
\path[every node/.style={font=\sffamily\small}]
(1) edge [] node[] {3} (5);
\path[every node/.style={font=\sffamily\small}]
(1) edge [] node[] {3} (6);
\path[every node/.style={font=\sffamily\small}]
(1) edge [] node[] {3} (7);
\path[every node/.style={font=\sffamily\small}]
(1) edge [] node[] {3} (8);
\path[every node/.style={font=\sffamily\small}]
(2) edge [] node[] {3} (3);
\path[every node/.style={font=\sffamily\small}]
(2) edge [] node[] {3} (4);
\path[every node/.style={font=\sffamily\small}]
(2) edge [] node[] {3} (5);
\path[every node/.style={font=\sffamily\small}]
(2) edge [] node[] {3} (6);
\path[every node/.style={font=\sffamily\small}]
(2) edge [] node[] {3} (7);
\path[every node/.style={font=\sffamily\small}]
(2) edge [] node[] {3} (8);
\path[every node/.style={font=\sffamily\small}]
(3) edge [] node[] {3} (4);
\path[every node/.style={font=\sffamily\small}]
(3) edge [] node[] {3} (5);
\path[every node/.style={font=\sffamily\small}]
(3) edge [] node[] {3} (6);
\path[every node/.style={font=\sffamily\small}]
(3) edge [] node[] {3} (7);
\path[every node/.style={font=\sffamily\small}]
(3) edge [] node[] {3} (8);
\path[every node/.style={font=\sffamily\small}]
(4) edge [] node[] {3} (5);
\path[every node/.style={font=\sffamily\small}]
(4) edge [] node[] {3} (6);
\path[every node/.style={font=\sffamily\small}]
(4) edge [] node[] {3} (7);
\path[every node/.style={font=\sffamily\small}]
(4) edge [] node[] {3} (8);
\path[every node/.style={font=\sffamily\small}]
(5) edge [] node[] {3} (6);
\path[every node/.style={font=\sffamily\small}]
(5) edge [] node[] {3} (7);
\path[every node/.style={font=\sffamily\small}]
(5) edge [] node[] {3} (8);
\path[every node/.style={font=\sffamily\small}]
(6) edge [] node[] {3} (7);
\path[every node/.style={font=\sffamily\small}]
(6) edge [] node[] {3} (8);
\path[every node/.style={font=\sffamily\small}]
(7) edge [] node[] {3} (8);
\end{tikzpicture}  $$

\begin{theorem} \label{T3.10}
Consider $ \rho_G$ as the density operator of a weighted graph $(G,a)$ with $n=2^m$ vertices where $(G,a)$ represents an $n$-qubit quantum state. If the reduced density operator ${\rho_G}^A$ is singular, then the quantum discord of the quantum state is zero, i.e., $D_Q(G,a)=0$.

\end{theorem}
Proof:Consider $\rho_G =[\rho_{ij}]_{2^m \times 2^m}$ as a density operator of a weighted graph $(G,a)$ with $n=2^m$ vertices where $(G,a)$ represents an $n$-qubit quantum state.

Since $det(\rho^A_G) = 0$, we have
\begin{equation*}
(\rho_{11}+\hdots+\rho_{\frac{n}{2}\frac{n}{2}})(\rho_{{\frac{n}{2}+1}{\frac{n}{2}+1}}+\hdots+\rho_{nn}) ={\abs{(\rho_{1{\frac{n}{2}+1}}+\rho_{2{\frac{n}{2}+2}}+\hdots+\rho_{\frac{n}{2}n})}}^2.
\end{equation*}

For $det(L^A)=0$, it is established that the density operator can be expressed as a product state \cite{joshi2022entanglement}, where each block commutes with one another. Therefore, it has been demonstrated.

			\section{Conclusion} \label{4}

   In this article, we presented a comprehensive analysis to study and characterize the properties of weighted graphs. The focus of our study revolved around the manifestation of zero discord as we presented significant findings in this area. Using a systematic approach, we divided the density matrix into distinct blocks, allowing for a thorough evaluation of zero discord across various quantum states. The outcomes of our investigation will have important implications for foundational aspects of quantum communication and computing.

\bibliographystyle{plain}
\bibliography{Anoopa_Discord}

\begin{thebibliography}{10}

\bibitem{barnett2009quantum}
Stephen Barnett.
\newblock {\em Quantum information}, volume~16.
\newblock Oxford University Press, 2009.

\bibitem{bennett1996concentrating}
Charles~H Bennett, Herbert~J Bernstein, Sandu Popescu, and Benjamin Schumacher.
\newblock Concentrating partial entanglement by local operations.
\newblock {\em Physical Review A}, 53(4):2046, 1996.

\bibitem{bennett1999quantum}
Charles~H Bennett, David~P DiVincenzo, Christopher~A Fuchs, Tal Mor, Eric
  Rains, Peter~W Shor, John~A Smolin, and William~K Wootters.
\newblock Quantum nonlocality without entanglement.
\newblock {\em Physical Review A}, 59(2):1070, 1999.

\bibitem{bennett1996mixed}
Charles~H Bennett, David~P DiVincenzo, John~A Smolin, and William~K Wootters.
\newblock Mixed-state entanglement and quantum error correction.
\newblock {\em Physical Review A}, 54(5):3824, 1996.

\bibitem{braunstein2006laplacian}
Samuel~L Braunstein, Sibasish Ghosh, and Simone Severini.
\newblock The laplacian of a graph as a density matrix: a basic combinatorial
  approach to separability of mixed states.
\newblock {\em Annals of Combinatorics}, 10(3):291--317, 2006.

\bibitem{dakic2010necessary}
Borivoje Daki{\'c}, Vlatko Vedral, and {\v{C}}aslav Brukner.
\newblock Necessary and sufficient condition for nonzero quantum discord.
\newblock {\em Physical review letters}, 105(19):190502, 2010.

\bibitem{datta2008quantum}
Animesh Datta, Anil Shaji, and Carlton~M Caves.
\newblock Quantum discord and the power of one qubit.
\newblock {\em Physical review letters}, 100(5):050502, 2008.

\bibitem{deutsch1996quantum}
David Deutsch, Artur Ekert, Richard Jozsa, Chiara Macchiavello, Sandu Popescu,
  and Anna Sanpera.
\newblock Quantum privacy amplification and the security of quantum
  cryptography over noisy channels.
\newblock {\em Physical review letters}, 77(13):2818, 1996.

\bibitem{divincenzo1995two}
David~P DiVincenzo.
\newblock Two-bit gates are universal for quantum computation.
\newblock {\em Physical Review A}, 51(2):1015, 1995.

\bibitem{dur2003multiparticle}
Wolfgang D{\"u}r, Hans Aschauer, and H-J Briegel.
\newblock Multiparticle entanglement purification for graph states.
\newblock {\em Physical review letters}, 91(10):107903, 2003.

\bibitem{fanchini2010non}
FF~Fanchini, T~Werlang, CA~Brasil, LGE Arruda, and AO~Caldeira.
\newblock Non-markovian dynamics of quantum discord.
\newblock {\em Physical Review A}, 81(5):052107, 2010.

\bibitem{gu2021fast}
Xiu Gu, Jorge Fern{\'a}ndez-Pend{\'a}s, Pontus Vikst{\aa}l, Tahereh Abad,
  Christopher Warren, Andreas Bengtsson, Giovanna Tancredi, Vitaly Shumeiko,
  Jonas Bylander, G{\"o}ran Johansson, et~al.
\newblock Fast multiqubit gates through simultaneous two-qubit gates.
\newblock {\em PRX Quantum}, 2(4):040348, 2021.

\bibitem{hassan2007combinatorial}
Ali Saif~M Hassan and Pramod~S Joag.
\newblock A combinatorial approach to multipartite quantum systems: basic
  formulation.
\newblock {\em Journal of Physics A: Mathematical and Theoretical},
  40(33):10251, 2007.

\bibitem{horn2012matrix}
Roger~A Horn and Charles~R Johnson.
\newblock {\em Matrix analysis}.
\newblock Cambridge university press, 2012.

\bibitem{horodecki2001separability}
Micha{\l} Horodecki, Pawe{\l} Horodecki, and Ryszard Horodecki.
\newblock Separability of n-particle mixed states: necessary and sufficient
  conditions in terms of linear maps.
\newblock {\em Physics Letters A}, 283(1-2):1--7, 2001.

\bibitem{huang2011new}
Jie-Hui Huang, Lei Wang, and Shi-Yao Zhu.
\newblock A new criterion for zero quantum discord.
\newblock {\em New Journal of Physics}, 13(6):063045, 2011.

\bibitem{joshi2022entanglement}
Anoopa Joshi, Parvinder Singh, and Atul Kumar.
\newblock Entanglement and separability of graph laplacian quantum states.
\newblock {\em Quantum Information Processing}, 21(4):152, 2022.

\bibitem{joshi2024separability}
Anoopa Joshi, Parvinder Singh, and Atul Kumar.
\newblock Separability of graph laplacian quantum states: Utilizing unitary
  operators, neighbourhood sets and equivalence relation.
\newblock {\em arXiv preprint arXiv:2401.02289}, 2024.

\bibitem{joshi2018concurrence}
Anoopa Joshi, Ranveer Singh, and Atul Kumar.
\newblock Concurrence and three-tangle of the graph.
\newblock {\em Quantum Information Processing}, 17(12):327, 2018.

\bibitem{knill1998power}
Emanuel Knill and Raymond Laflamme.
\newblock Power of one bit of quantum information.
\newblock {\em Physical Review Letters}, 81(25):5672, 1998.

\bibitem{lanyon2008experimental}
Ben~P Lanyon, Marco Barbieri, Marcelo~P Almeida, and Andrew~G White.
\newblock Experimental quantum computing without entanglement.
\newblock {\em Physical review letters}, 101(20):200501, 2008.

\bibitem{luo2016entanglement}
Shunlong Luo.
\newblock Entanglement as minimal discord over state extensions.
\newblock {\em Physical Review A}, 94(3):032129, 2016.

\bibitem{modi2012classical}
Kavan Modi, Aharon Brodutch, Hugo Cable, Tomasz Paterek, and Vlatko Vedral.
\newblock The classical-quantum boundary for correlations: Discord and related
  measures.
\newblock {\em Reviews of Modern Physics}, 84(4):1655, 2012.

\bibitem{nielsen2000quantum}
Michael~A Nielsen and Isaac~L Chuang.
\newblock Quantum computation and quantum information, 2000.

\bibitem{nielsen2010quantum}
Michael~A Nielsen and Isaac~L Chuang.
\newblock {\em Quantum computation and quantum information}.
\newblock Cambridge university press, 2010.

\bibitem{ollivier2001quantum}
Harold Ollivier and Wojciech~H Zurek.
\newblock Quantum discord: a measure of the quantumness of correlations.
\newblock {\em Physical review letters}, 88(1):017901, 2001.

\bibitem{piani2008no}
Marco Piani, Pawe{\l} Horodecki, and Ryszard Horodecki.
\newblock No-local-broadcasting theorem for multipartite quantum correlations.
\newblock {\em Physical review letters}, 100(9):090502, 2008.

\bibitem{rulli2011global}
CC~Rulli and MS~Sarandy.
\newblock Global quantum discord in multipartite systems.
\newblock {\em Physical Review A}, 84(4):042109, 2011.

\bibitem{shabani2009vanishing}
Alireza Shabani and Daniel~A Lidar.
\newblock Vanishing quantum discord is necessary and sufficient for completely
  positive maps.
\newblock {\em Physical review letters}, 102(10):100402, 2009.

\bibitem{wang2010non}
Bo~Wang, Zhen-Yu Xu, Ze-Qian Chen, and Mang Feng.
\newblock Non-markovian effect on the quantum discord.
\newblock {\em Physical Review A}, 81(1):014101, 2010.

\bibitem{wang1993some}
Bo-Ying Wang and Ming-Peng Gong.
\newblock Some eigenvalue inequalities for positive semidefinite matrix power
  products.
\newblock {\em Linear Algebra and Its Applications}, 184:249--260, 1993.

\bibitem{werlang2009robustness}
T~Werlang, S~Souza, FF~Fanchini, and CJ~Villas Boas.
\newblock Robustness of quantum discord to sudden death.
\newblock {\em Physical Review A}, 80(2):024103, 2009.

\bibitem{wootters1998entanglement}
William~K Wootters.
\newblock Entanglement of formation of an arbitrary state of two qubits.
\newblock {\em Physical Review Letters}, 80(10):2245, 1998.

\bibitem{wootters2001entanglement}
William~K Wootters.
\newblock Entanglement of formation and concurrence.
\newblock {\em Quantum Information \& Computation}, 1(1):27--44, 2001.

\bibitem{yao2012geometric}
Yao Yao, Hong-Wei Li, Zhen-Qiang Yin, and Zheng-Fu Han.
\newblock Geometric interpretation of the geometric discord.
\newblock {\em Physics Letters A}, 376(4):358--364, 2012.

\end{thebibliography}

\end{document}